\documentclass[aps,prl,twocolumn,superscriptaddress,showpacs]{revtex4-1}
 \usepackage{graphicx}
\usepackage{afterpage} \usepackage{dcolumn} \usepackage{bm}\usepackage{amssymb}\usepackage{tabularx}

\begin{document}

\preprint{APS/123-QED}

\title{Tunable localization in an \em s\em-electron Kondo system at room temperature}

\author{J.L.M.~van~Mechelen}
\email{dook.vanmechelen@ch.abb.com}
\affiliation{ABB Corporate Research, Segelhofstrasse 1K, 5405 Baden-D\"attwil, Switzerland}

\affiliation{Department of Physics and Astronomy, VU University, De Boelelaan 1081, 1081 HV Amsterdam, The Netherlands}
\author{M.J.~van~Setten}
\email{present affiliation: IMEC, Kapeldreef 75, 3001 Heverlee, Belgium}
\affiliation{Nanoscopic Physics, Institute of Condensed Matter and Nanosciences, Universit\'e catholique de Louvain, 1348 Louvain-la-Neuve, Belgium}

\date{\today}
\maketitle

{\bf
To achieve room-temperature superconductivity, a mechanism is needed that provides heavy quasiparticles at room temperature.\cite{Qazilbash2009,Basov2011a} In heavy fermion systems such localization is prototypically present only at liquid helium temperatures. In these $f$-electron Kondo systems, conduction electrons magnetically couple to localized moments, enhancing their mass and scatting time. These quasiparticles may form Cooper pairs and cause unconventional superconductivity with a critical temperature $T_c$ of the order of the Fermi energy $\varepsilon_F$.\cite{Khomskii2014} In relative terms, this $T_c$ is much larger than in cuprate or BCS superconductors for which $T_c\ll \varepsilon_F$.\cite{Khomskii2014} This suggests that Kondo systems in general have the potential to be high-temperature superconductors.\cite{Pines2013} For this to occur, strong correlations that cause electron localization need to take place at much larger temperatures. Here we show that metal hydrides manifest strong electron correlations in a single $3d_{x^2-y^2}$ band at the Fermi level, similar to the cuprates but at room temperature. Hole doping of this band, by varying the hydrogen content, causes divergence of the carrier mass and suggests the approach of an ordered Mott transition with signatures of a correlated metallic Kondo lattice. These room-temperature phenomena are expected to be widespread across hydrogen-rich compounds, and offer a promising novel ground to encounter unconventional superconductivity in the class of the metallic hydrides.} 

Dense metallic hydrides form a material class with a tremendous societal potential for high-temperature superconductivity and energy storage.\cite{Ashcroft2004,Mohtadi2016} Their physics, however, is far from understood. Adding hydrogen to a metallic host strongly influences the electronic and structural properties. It may bind to impurities, manifests itself as an inert lattice gas and, at larger concentrations, can induce a metal-to-insulator transition.\cite{Huiberts1996} Dense hydrides, moreover, manifest high energy lattice vibrations and strong electron-phonon coupling and are therefore seen as an ideal testbed for conventional, electron-phonon mediated, superconductivity at high temperatures.\cite{Ashcroft1968,Ashcroft2004} Experimental confirmation was indeed obtained, first in elemental hydrides such as PdH$_x$ with $T_c\sim10$\,K and more recently in SH$_3$ at $T_c\sim203$\,K although under gigantic pressures.\cite{PdHsupercond,Drozdov2015}

The path to even larger $T_c$ in metallic hydrides, without hydrostatic pressure, may be offered in a slightly different direction. In most high-temperature superconductors Cooper pair formation is caused by strong electron-electron correlations close to a Mott insulating state, rather than the electron-phonon interaction. Close to the Mott metal-to-insulator (MI) transition Kondo physics plays an important role,\cite{Georges1996} and is expected to strongly enhance $T_c$.\cite{Capone2002} In rare earth hydrides, the puzzling MI transition, which occurs at room temperature at a non-stoichiometric hydrogen concentration, cannot be explained with simple band theory.\cite{Huiberts1996,Eder1997} It has been proposed that additional hydrogen atoms added to the metallic $d^1$ dihydrides LaH$_2$ and YH$_2$ can be seen as Kondo impurities that antiferromagnetically couple to charge carriers,  eventually leading to a Mott insulating state.\cite{Eder1997, Ng1999, Ng1997} This strongly correlated state may involve $s$-$d$ hybridization, which is much larger than the $p$-$d$ hybridization in oxide and pnictide superconductors. If such a Kondo lattice of H impurities exists, it is therefore expected to have a Kondo temperature $T_K>10^3$\,K.\cite{Takada2015} This is an order of magnitude larger than $T_K$ of the iron pnictides superconductors, which are $d$-electron Kondo systems.\cite{Wu2016} Given that for most systems $T_K\propto T_c$, this implies that hydrogen-based Kondo systems are a prosperous candidate for high-temperature superconductivity.

\begin{figure*}
\begin{center}
\includegraphics [width=\linewidth]{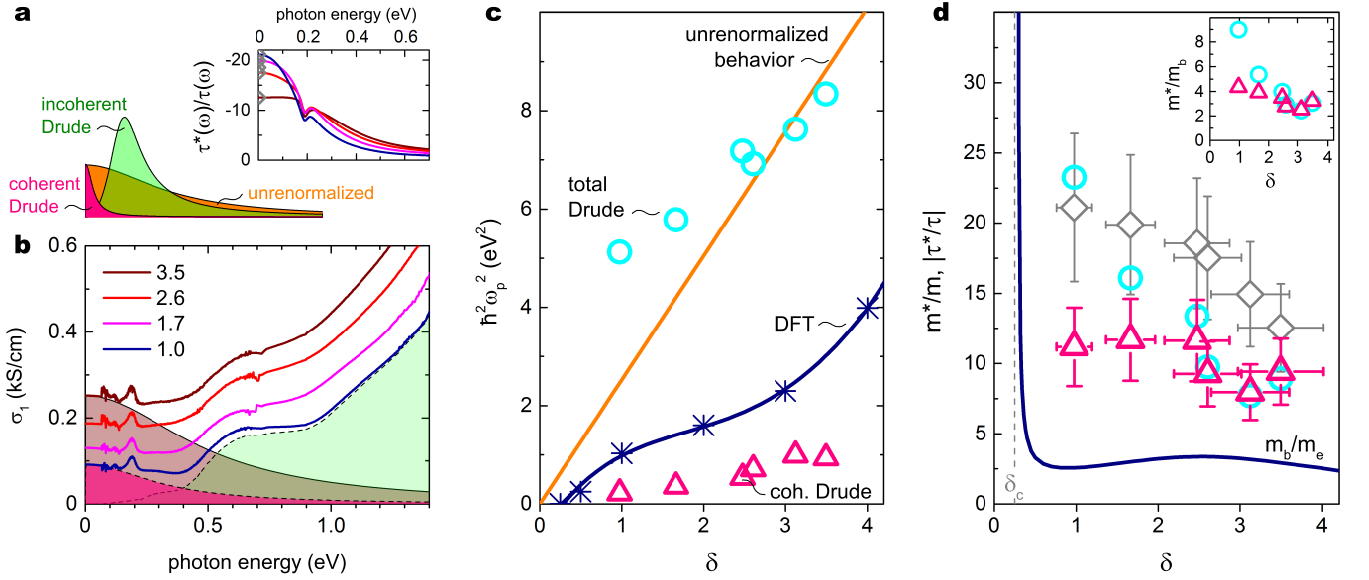}
\caption{
(a) Illustration of the Drude weight of a renormalized system that is divided into a coherent an incoherent part, $W_{\text{coh}}$ and $W_{\text{incoh}}$, respectively, as compared to an unrenormalized system. The inset shows the experimental scattering rate renormalization $\tau^*/\tau$. (b) Experimental $\sigma_1(\omega)$ of Mg$_y$NiH$_{4-\delta/8}$ in the infrared range for indicated values of $\delta$ at 298\,K. $W_{\text{coh}}$ and $W_{\text{incoh}}$ are shown by surfaces, colored as in panel (a). Peaks around 0.08 eV and 0.19 eV are Ni--H bending and stretching vibration modes, respectively. (c) $W_{\text{coh}}$ from optical data (triangles) and from DFT calculations (crosses, line is a guide to the eye) is compared to unrenormalized behavior, where each charge carrier is itinerant (orange line). Also shown is $W_{\text{tot}}=W_{\text{coh}}+W_{\text{incoh}}$ from optical data for selected values of $\delta$ (circles, see Supplementary Information). (d) Mass renormalization $m^*/m_e=W_{\text{tot}}/W_{\text{coh}}$ from optical data (circles), $m^*/m_e$ with $W_{\text{tot}}$ from Hall data and $W_{\text{coh}}$ from optical data (triangles), $\tau^*/\tau$ at $\omega=0$ (lozenges) as in panel (a), and the band mass $m_b/m_e$ from DFT (solid line). The inset shows $m^*/m_b$. } \label{fig:mass}
\end{center}
\end{figure*}

Transition metal hydrides also exhibit MI transitions around hydrogen concentrations unlikely to cause a simple metal to band-insulator transition. Here we study the MI transition in the complex metal hydride Mg$_2$NiH$_x$ at 298\,K. Removal of H from the band insulator Mg$_2$NiH$_4$ breaks H--Ni bonding states and effectively hole dopes this $d^{10}$ system in the top of the Ni $3d_{x^2-y^2}$ band. We use optical spectroscopy to determine the charge dynamics within this band of single phase Mg$_y$NiH$_x$ thin films. The charge carrier density $\delta=32-8x$ is accurately tuned in the range $1.0\leqslant\delta\leqslant3.5$ by a slight variation of the Mg/Ni ratio, $1.45\leqslant y\leqslant1.52$, which has a strong effect on the hydrogenation kinetics and thus on $\delta$ (see Supplementary Information). This range of $x$ is far beyond the structural phase transition from hexagonal ($P6_2 22$) to monoclinic ($C2/c$) occurring for $1<x< 3$,\cite{Zolliker1986,Lohstroh2004} and away from the band insulator phase at $x=4$. For such a system, itinerant electron behavior would cause a Drude peak in the optical conductivity $\sigma_1(\omega)$ centered at zero photon frequency $\omega$. Carrier localization, however, would split the Drude peak in a renormalized coherent part at $\omega=0$ and an incoherent part at infrared frequencies (Fig.~\ref{fig:mass}a).\cite{VanMechelen2008}

The key result of our optical study is that metallic Mg$_y$NiH$_{4-\delta/8}$ manifests a coherent Drude peak with weight $W_{\text{coh}}=\hbar^2\omega_p^2$ that is substantially less than expected for unrenormalized itinerant electron behavior, which directly points to localization at room temperature (Fig.~\ref{fig:mass}b,c). The lacking weight is recovered in a broad infrared band,  shown by $\sigma_1(\omega)$, which forms the incoherent part of the Drude weight $W_{\text{incoh}}$ (Fig.~\ref{fig:mass}b). The degree of localization is expressed by the effective electron mass $m^*/m_e=W_{\text{tot}}/W_{\text{coh}}$, where $m_e$ is the free electron mass. We obtain the total charge carrier weight $W_{\text{tot}}$ in two ways: by taking the sum of the coherent and incoherent Drude weights, and from Hall data through $W_{\text{tot}}=\delta\hbar^2 e^2/\epsilon_0 m_eV$, with $V$ the volume of the unit cell.\cite{Enache2004} The two approaches both evince a lack of $W_{\text{coh}}$ and thus a significant mass enhancement, which increases towards the insulating phase (Fig.~\ref{fig:mass}d).

In strongly correlated systems, localization enhances both the electron mass and, because of the decreased Fermi velocity, also the scattering time, as mentioned before. That is, quasiparticle renormalization affects both $m^*$ and the scattering rate $1/\tau^*(\omega)$, such that the effect cancels at $\omega=0$.\cite{Millis1987} We determine $\tau^*(0)/\tau(0)$ from an extended Drude analysis (Fig.~\ref{fig:mass}a, Supplementary Information). $\tau^*(0)/\tau(0)$ displays a strong electron renormalization that has a similar behavior as $m^*/m_e$, thereby pointing to an enhanced scattering time for heavier electrons (Fig.~\ref{fig:mass}d).

\begin{figure*}
\includegraphics [width=\linewidth]{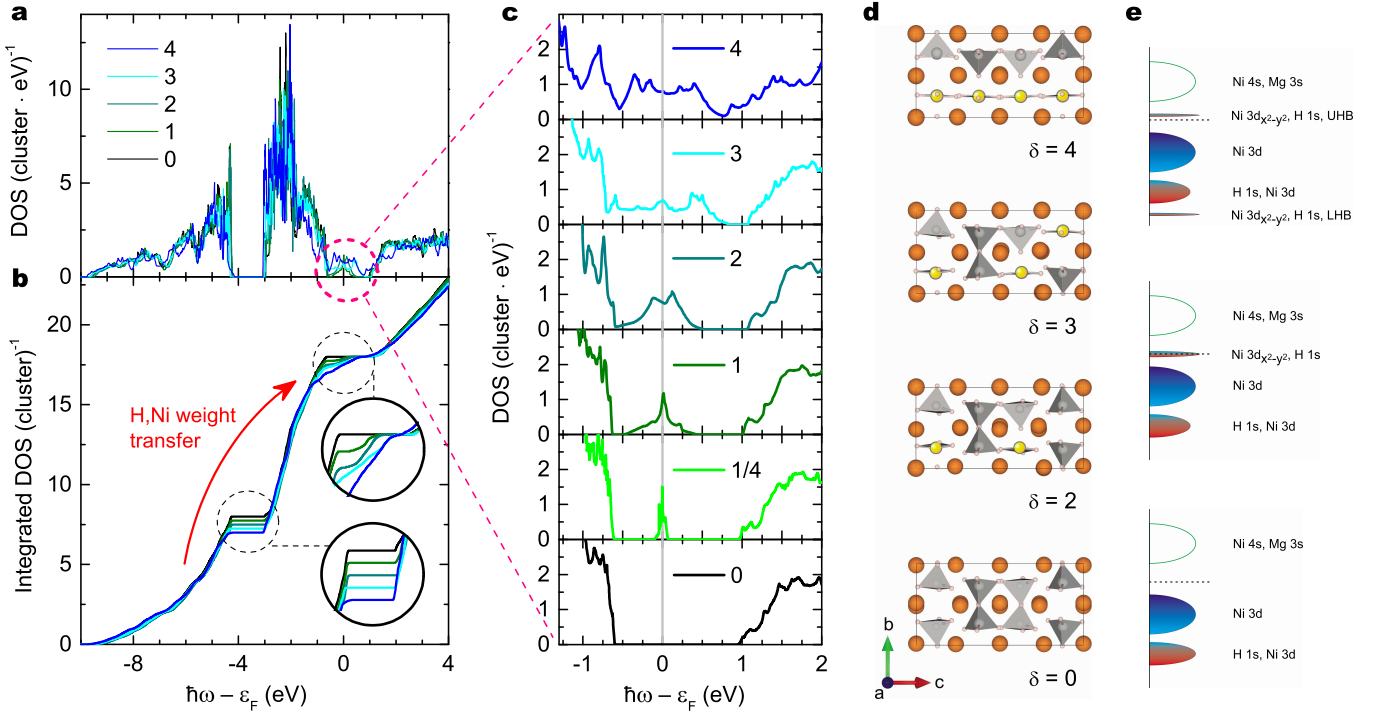}
\caption{(a) Electronic DOS of Mg$_2$NiH$_{4-\delta/8}$ for indicated values of $\delta$ calculated using DFT; (b) Integrated DOS showing the transfer of hybridized H and Ni spectral weight to the states around $\varepsilon_F$. (c) DOS as in (a) enlarged around $\varepsilon_F$. (d) Mg$_2$NiH$_{4-\delta/8}$ unit cell (orange: Mg, yellow: Ni of NiH$_3$ clusters, gray: Ni of NiH$_4$, pink: H). (e) Cartoon representation of the band structure of band insulating Mg$_2$NiH$_4$ (bottom), metallic Mg$_2$NiH$_{4-\delta/8}$ as calculated from DFT (middle), and highly correlated Mg$_2$NiH$_{4-\delta/8}$ as suggested by the experimental data (top).} \label{fig:dos}
\end{figure*}

In order to estimate the influence of band effects on the charge carrier renormalization, which are not taken into account in $m^*/m_e$, we performed Density Functional Theory (DFT) calculations within the generalized gradient approximation (GGA). The density of states (DOS) of Mg$_2$NiH$_4$ shows to be a band insulator with Ni $3d$ valence bands below $\varepsilon_F$ and empty Mg $3s$ and Ni $4s$ bands about 1.65~eV above these bands (Fig.~\ref{fig:dos}a,c), consistent with earlier reports.\cite{Garcia1999} This indicates that the charge distribution in Mg$_2$NiH$_4$ is governed by the strong electronegativities of H and Ni. Ionic bonding occurs between Mg$^{2+}$ and the complex NiH$_4^{4-}$. Strong covalent bonding between Ni and H causes deep lying H bonding bands ($-10\ \text{eV}<\hbar\omega-\varepsilon_F<-4\ \text{eV}$), which indicates H to be in the anionic $1s^2$ state.\cite{Garcia1999} Therefore, Ni is in the particular $3d^{10}4s^0$ state and all its $3d$ bands are below $\varepsilon_F$. Calculations of Mg$_y$NiH$_{4-\delta/8}$  super cells with $\delta$ H vacancies (Fig. 2d) --- no more than one per NiH$_4$ cluster --- show that each H vacancy dopes the system with one charge carrier. For $\delta\leqslant1$, i.e., until one H vacancy per Mg$_{16}$Ni$_8$H$_{32}$ unit cell, we find that a narrow half-filled band occurs around $\varepsilon_F$. These in-gap states have Ni $3d_{x^2-y^2}$ character hybridized with H $1s$ owing to the vacancy state. This is confirmed by the integrated DOS that shows a complete spectral weight transfer from the Ni-H bonding bands to the top of the in-gap states (see Fig.~\ref{fig:dos}a,b). The removal of one H from a tetrahedral NiH$_4$ cluster changes its structure to square-planar NiH$_3$, thereby minimizing the electrostatic energy. Due to the increased Ni-H overlap this pushes the Ni $3d_{x^2-y^2}$ band above $\varepsilon_F$, similar as reported for the  Mg$_2$NiH$_4$ model system with square planar NiH$_4$ clusters.\cite{Garcia1999} Furthermore, removal of an additional H (on a different cluster) creates a similar, but due to the monoclinic structure distinct, in-gap state (cf. Fig.~\ref{fig:dos}c for $\delta>1$). A slight variation of $y$, comparable to the sample compositions, qualitatively does not alter the DOS (see Supplementary Information).

The coherent Drude weight determined from the DFT calculations, $W_{\text{DFT}}=\hbar^2\omega_{p, \text{DFT}}^2$, is proportional to $\delta$ and becomes close to zero for $\delta_c\approx0.25$ (Fig.~\ref{fig:mass}c). At this concentration the H impurities are too distant to cause significant dispersion and sustain dc conductivity. Consequently, the band mass $m_b/m_e=W_{\text{tot}}/W_{\text{DFT}}$ diverges at this point. For $\delta>0.5$, $m_b/m_e\approx3$ almost independent of the carrier concentration. The experimental optical mass $2<m^*/m_b<9$ is thus also strongly renormalized when taking band effects into account.

Both our experiments and DFT calculations combined unanimously show, by the properties of the electron mass and scattering rate, strong electron correlations in the single occupied $d_{x^2-y^2}$ in-gap states. These strong correlations in turn also renormalize the width of the in-gap states $\Gamma$. Furthermore, since the on-site Coulomb interaction on the Ni atom $U\sim10$\,eV is much larger than the unrenormalized in-gap band width, $\Gamma_K\sim0.4$\,eV for $\delta=1$, these states will spit in a lower and upper Hubbard band, and a Kondo resonance peak at $\varepsilon_F$ hosting the localized carriers (Fig. 2e). The presence of Kondo physics close to a Mott transition, although generally present,\cite{Georges1996} is here even more intuitive because of the impurity character of the hydrogen vacancies that antiferromagnetically couple to the Ni conduction electrons below $T_K$. $k_BT_K$ is equal to the width of the Kondo resonance $\gamma_K\lesssim\Gamma_K$, i.e., of the order $10^3$\,K, and thus inside the predicted ballpark.\cite{Takada2015} This is further confirmed by the periodic Anderson model, for which $m^*/m_b=(\Delta_{\text{dir}}/T_K)^2$, where $\Delta_{\text{dir}}\approx0.4\,$eV is the direct hybridization gap energy (see Fig.~\ref{fig:mass}b) with $m^*/m_b$ as in the inset of Fig.~\ref{fig:mass}d. Below $T_K$ all magnetic moments are screened and the material becomes non-magnetic. Indeed, the magnetic susceptibility of Mg$_2$NiH$_x$ has been reported to show a pronounced decrease upon $\delta\rightarrow1.6^+$ at room temperature.\cite{Stucki1980} Mg$_2$NiH$_{4-\delta/8}$ thus crosses a Mott metal-to-insulating transition well before $\delta=0$.

\begin{figure}
\includegraphics [width=\linewidth]{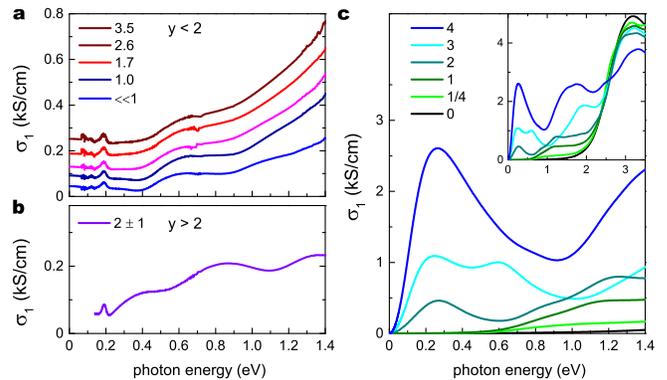}
\caption{(a) Experimental $\sigma_1(\omega)$ at 298\,K for indicated values of $\delta$ and $y<2$ (as in Fig. 1b). (b) Experimental $\sigma_1(\omega)$ at 298\,K for $y>2$. (c) Calculated interband contribution to $\sigma_1(\omega)$ using DFT for $y=2$ and $\delta$ indicated in the legend. The inset shows $\sigma_1(\omega)$ in an enlarged range.} \label{fig:s1}
\end{figure}

A Kondo lattice occurs when the observed Kondo phenomena occur periodically. A close inspection of the optical data indeed hints to a Kondo impurity lattice. For all compositions, $\sigma_1(\omega)$ shows typical characteristics of an ordered (poly)crystalline material opposite to that of an amorphous electron glass, such as a coherent Drude peak, well-defined phonon modes and a clearly distinct band gap. Particularly for $\delta\geq2$ and $y\gtrsim2$, we observe an absorption band around 0.3\,eV as in the, periodic, DFT results for this composition (Fig.~\ref{fig:s1}b,c). Moreover, the resemblance between the DFT results and the optical measurements can further be read off above 1.5\,eV, where both have a very similar trend upon changing the Mg/Ni ratio (see Supplementary Information). For these reasons, we conjecture that in the vicinity of the Mott transition, Mg$_2$NiH$_x$ has signatures of an ordered Kondo lattice consisting of vacancy states with $s$-character that localize Ni $d$ carriers.

Although strong electron correlations have never been experimentally observed in hydrides, Mott insulating phases and concomitant mass divergences have been reported for other systems such as Si:P, $^3$He and Sr doped LaTiO$_3$.\cite{Rosenbaum1983,Tokura1993,Casey2003} Among those, Si:P may appear closest to the current hydride, but the disordered arrangement of P in the Si lattice rather causes a disorder induced Mott transition, opposite to the ordered Mott transition in Mg$_2$NiH$_x$. The main difference with all known systems that manifest a pure Mott phase, is that this state of matter in Mg$_2$NiH$_x$ occurs at room temperature.

The strong correlations observed in hole-doped Mg$_2$NiH$_x$ are likely to be omnipresent in related hydride systems. Previous work shows that systems similar to Mg$_2$NiH$_4$ and Mg$_2$FeH$_6$, with Ni or Fe substituted by Mn, Ni, Fe, Co and Cu have comparable electronic structures to Mg$_2$NiH$_{4-\delta/8}$.\cite{Batalovic2014, VanSetten2007a} Calculations on these systems evince an increased density of states similar to the in-gap states of Mg$_2$NiH$_{4-\delta/8}$, which is not located mid-gap, but close to either the conduction or valence band. Besides, also the rare-earth hydrides YH$_x$ and LaH$_x$ are expected to possess strong electron correlations.\cite{Eder1997,Ng1997} It is further interesting noticing that DMFT calculations of H impurities in Ge$_8$ and Ge$_{64}$ supercells show a split H and Ge hybridized Hubbard band close to $\varepsilon_F$,\cite{Sims2013} which maybe implies that the class of strongly correlated hydrides is even larger than ones that show a MI transition.

On the question how encounter superconductivity in Mg$_y$NiH$_x$, we conjecture that maintaining stoichiometry for $y$ is crucial as our work showed that only for $y\gtrsim2$ an ordered Kondo lattice forms. The hydrogen concentration $x$ is most promising where strong renormalization effects take place, such as for $1\leqslant\delta\leqslant2$, forming a correlated metal with $m_b/m^*\sim0.2$ in the ballpark to allow superconductivity.\cite{Qazilbash2009} Above all, a comprehensive study is favored employing experimental techniques that directly evince a superconducting state.\\

\scriptsize {\bf Acknowledgements} We thank S.~Biermann, W.E.~Pickett, D.~van der Marel and G.A.~de~Wijs for insightful discussions. We are grateful to R.~Griessen and B.~Dam for providing the experimental facilities, and to W.~Lustenhouwer, S.~Kars, H.~Schreuders and J.H.~Rector for technical support. Financial support by the Netherlands Organization for Scientific Research (NWO) is gratefully acknowledged.\\

\scriptsize {\bf Author Contributions}\ \ J.L.M.v.M. proposed and lead the study, conducted the optical and structural measurements, analyzed and interpreted the data.
M.J.v.S. performed the DFT calculations and their interpretation in the context of the experimental data. Both authors wrote the manuscript.\\

\scriptsize {\bf Data availability} The datasets generated and analysed during the current study are available from the corresponding author on reasonable request.

\end{document}

\title{\em Suppementary Information\em\\Tunable localization in an $s$-electron Kondo system at room temperature}
\author{J.L.M.~van~Mechelen}\email{dook.vanmechelen@ch.abb.com}
\affiliation{ABB Corporate Research, Segelhofstrasse 1K, 5405 Baden-D\"attwil, Switzerland}
\affiliation{Department of Physics and Astronomy, VU University, De Boelelaan 1081, 1081 HV Amsterdam, The Netherlands}
\author{M.J.~van~Setten}\email{michiel.vansetten@imec.be}\email{present affiliation: IMEC, Kapeldreef 75, 3001 Heverlee, Belgium}
\affiliation{Nanoscopic Physics, Institute of Condensed Matter and Nanosciences, Universit\'e catholique de Louvain, 1348 Louvain-la-Neuve, Belgium}

\date{\today}
\maketitle

\tableofcontents
\newpage

\renewcommand\thefigure{S\arabic{figure}}
\renewcommand{\theequation}{S\arabic{equation}}
\renewcommand{\thetable}{S\arabic{table}}

\section{Thin film growth, crystal structure and morphology}
\subsection{Thin film growth and composition determination}
\mgni thin films with a composition gradient are deposited at room temperature by dc/rf magnetron co-sputtering (ATC 2400, AJA International, USA). Within the sputtering chamber (base pressure $10^{-7}$ Pa), a Mg and Ni target are placed in sputtering guns that are oriented off-axis. Adjustment of the gun angle and applied power allowed deposition of Mg$_y$Ni thin films with $1.4\leq y\leq 5$  on optically polished substrates chosen according to the measurement technique (see Table~\ref{tab:substrates}). All samples have been covered by a $4-20$ nm Pd layer to prevent oxidation and facilitate H$_2$ dissociation. For optical measurements this layer was 4 nm. Typical deposition rates are: $1.2€"-3.5$ \AA/s for Mg (sputtering power 118 W at 212 V rf), $0.3-0.7$ \AA/s for Ni (sputtering power 21 W at 299 V dc) and 1.3 \AA/s for Pd (sputtering power 50 W at 318 V dc).

Local compositions have been determined by Rutherford Backscattering Spectrometry (RBS) using 2 MeV He$^+$ ions with a 1 mm$^2$ spot size, and wavelength dispersive X-ray diffraction (WDS) using a JEOL JXA-8800M Electron Microprobe. The film thickness has been measured in the as-deposited state using a stylus profilometer (Veeco Dektak$^3$, USA).

\begin{table}[b]\caption{Substrates used for Mg$_y$Ni thin film growth classified per measurement technique.}\label{tab:substrates}
\begin{tabularx}{0.6\linewidth}{lcl}
\\
\hline
Measurement technique & \ \qquad \ & Used substrate\\
\hline
optical spectroscopy  ($0.06-0.5$ eV) && 2 mm KBr    \\
optical spectroscopy  ($0.14-3.3$ eV) && 1 mm CaF$_2$    \\
RBS, WDS, SEM && amorphous C  \\
XRD, AFM, profilometry && Si, SiO$_2$, sapphire  \\
dc resistivity && SiO$_2$  \\

\hline\\
\end{tabularx}
\end{table}

\begin{figure}
\begin{center}
\includegraphics [width=0.9\linewidth]{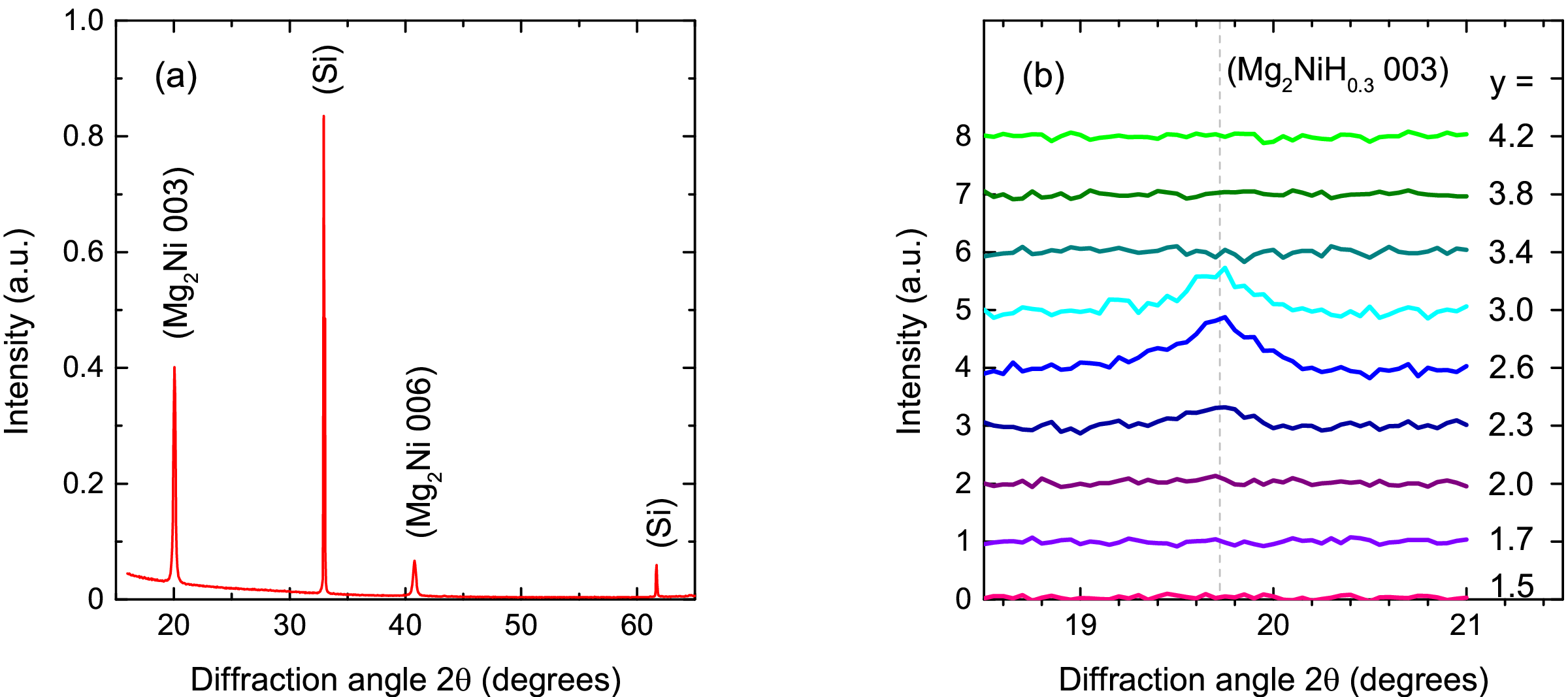}
\caption{X-ray diffraction pattern of (a) as-deposited Mg$_{y}$Ni for $y\approx2$ on a Si substrate at 298\,K showing the Mg$_2$Ni (003) and (006) reflections, and (b) hydrogenated Mg$_{y}$NiH$_{4-\delta/8}$ at $10^5$ Pa H$_2$ at 298\,K showing the solid-solution Mg$_2$NiH$_{0.3}$ (003) reflections (for clarity, the spectra are mutually shifted by one unit).} \label{fig:xrd}
\end{center}
\end{figure}

\subsection{Crystal structure}
Structural characterization is carried out by X-ray diffraction (XRD) using a Bruker D8 Discover with a $5-6$ mm Cu K$_{\alpha}$ beam with $\lambda=1.5418$\,\AA\ and mounted in a $\theta-2\theta$ geometry. The spectrometer was equipped with a slit detector and a Be dome to create an enclosed hydrogen atmosphere.

The diffracted pattern of as-deposited Mg$_y$Ni shows the (003) and (006) reflections of Mg$_2$Ni for $1.4\leq y\leq 4$, as well as those of the Si substrate (Fig.~\ref{fig:xrd}a). These reflections, and the absence of any traces of Ni or Mg, indicate that all our studied films are single phase crystalline Mg$_2$Ni.

Upon exposing the films to low hydrogen partial pressures of about 10$^2$ Pa, the (003) and (006) Mg$_y$Ni reflections occur at slightly lower diffraction angles, pointing to the formation of the solid-solution phase Mg$_2$NiH$_{0.3}$, in agreement with Ref.~\onlinecite{Lohstroh2004}. When the hydrogen partial pressure is increased to 10$^5$ Pa, at which the optical measurements are performed, the reflections of Mg$_2$NiH$_{0.3}$ vanish for $y<2.3$ and $y\geq3.4$, and strongly decrease to a trace amount level for $2.3\leq y\leq3.0$  (see Fig.~\ref{fig:xrd}). This result is in agreement with the earlier reported observation that in Mg$_{\approx2}$NiH$_{4-\delta/8}$ the Mg$_2$NiH$_{0.3}$ contribution vanishes for $\delta\leq8$.\cite{Lohstroh2004} Furthermore, also in this hydrogenated state, the x-ray data do not show any presence of (segregated) Ni or Mg for $y<2.3$, whereas $y>2.3$ displays traces of both Ni and Mg. Conform previous reports, coherent x-ray reflections of Mg$_2$NiH$_4$ cannot be observed.\cite{Lohstroh2004,Mongstad2012}.

Despite the absence of Mg$_2$NiH$_4$ reflections, we show here that our films are composed of single phase Mg$_y$NiH$_x$. Strong NiH$_4$ lattice modes, present at all studied Mg/Ni ratios around 0.08 and 0.19 eV, indicate that regular NiH$_x$ clusters are formed (see Figs.~1, \ref{fig:opticfit} and \ref{fig:s1}). Moreover, the presence of a Drude peak and a distinct band gap in the optical data implies that electronically the system shows (poly)crystalline behavior. In addition, we have calculated the optical properties of single phase Mg$_y$NiH$_x$, which describe our experimental data, and also previous reported data of Ref.~\onlinecite{Lohstroh2006}, to a high level of detail (see section \ref{sec:dft_y} and Fig.~\ref{fig:e2-visuv}). In order to exclude contamination of our samples with a second phase, we have tried to model our Mg$_y$NiH$_x$ optical data with an effective medium approach containing two phases, both for (i) Mg$_2$NiH$_4$ and MgH$_2$ and (ii) Mg$_2$NiH$_4$ and Ni. Both cases fail to describe the experimental data. The first case is similar and consistent with Ref.~\onlinecite{Lohstroh2006}. The second case predicts an infrared behavior which is far from what is experimentally observed, for all form factors and filling fractions of Ni inclusions. The reason is that a very small amount of metallic Ni inclusions drastically affects the infrared behavior of Mg$_y$NiH$_x$ close to the metal-to-insulator transition. From these observations, we conclude that our studied films with $1.0\leq\delta\leq 3.5$ are single phase Mg$_2$NiH$_{4-\delta/8}$.

\begin{figure}
\begin{center}
\includegraphics [width=\linewidth]{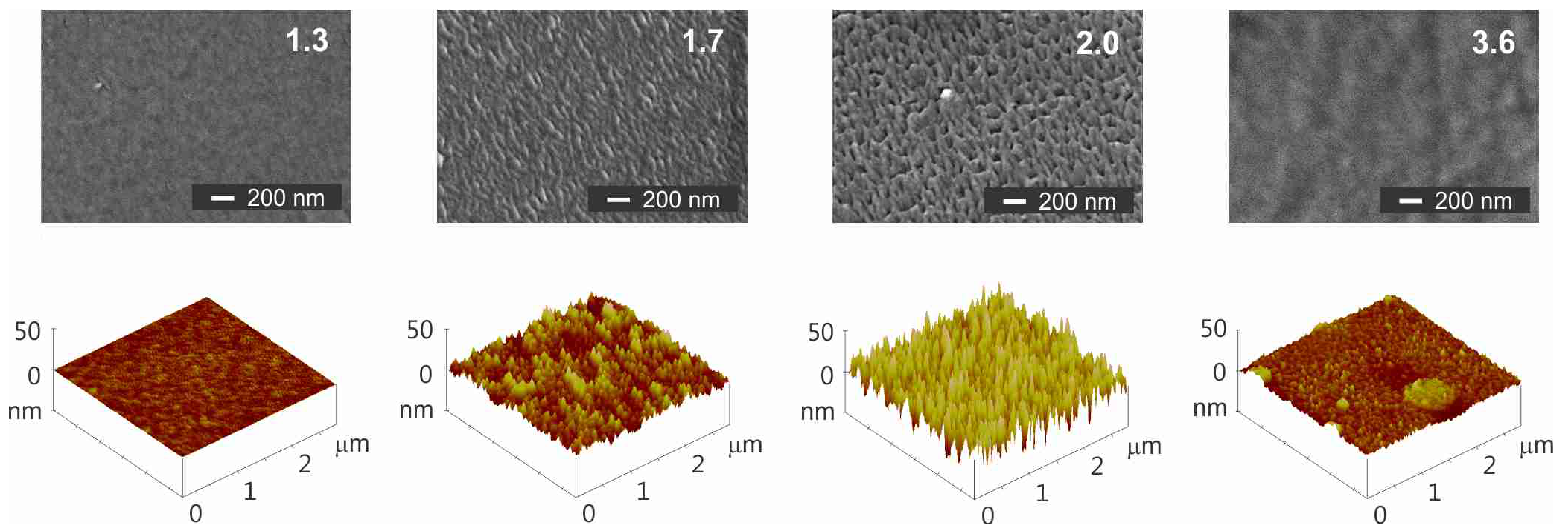}
\caption{Surface morphology of Mg$_y$Ni for indicated compositions $y$ determined by SEM (upper row) and AFM (bottom row). A strongly enhanced roughness can be observed around $y\approx2$.} \label{fig:morphology}
\end{center}
\end{figure}

\subsection{Morphology}
The surface morpholgy has been determined by atomic force microscopy (AFM) using a NanoScope III Atomic Force Microscope, and by scanning electron microscopy (SEM) using a JEOL JSM-6301F Scanning Electron Microscope. Fig.~\ref{fig:morphology} shows the surface morphology for selected values of $y$ probed by SEM at 4\,kV and by AFM in tapping mode using a silicon cantilever. Both techniques unanimously indicate that the gradient thin film is smooth on the nm scale well above and well below stoichiometry $y=2$. At stoichiometry, the roughness is strongly enhanced, which could be intrinsic, due to oxidation at the surface or due to an inhomogeneous distribution of Pd. RBS measurements on the same samples indicated that the amount of oxygen at the surface is composition independent and on average $1.74\times10^{16}\pm0.92\times10^{16}$ atoms/cm$^2$. Also the amount of Pd for all compositions is similar. The roughness further seems to scale with the amount of Mg$_2$Ni. We conjecture that the hydrogen kinetics depends on the surface morphology in a sense that a larger surface area enhances the hydrogen adsorption rate.

\section{Optical spectroscopy}\label{sec:optics}

\subsection{Experimental methods}
Optical spectroscopy is performed between 0.06 and 0.8~eV using a Bruker IFS 66/S FT-IR spectrometer with a liquid nitrogen cooled MCT detector, and between 0.72 and 3.3~eV using a Bruker IFS 66 FT-IR spectrometer with InGaAs and Si detectors. The spectrometers are equipped with a unit that allows quasi-simultaneous acquisition of transmission and reflection at near-normal incidence. Calibration for the incident spectrum has been done in transmission geometry using an aperture at the sample position. In this way, also reflectivity is obtained without reference sample. A hydrogen loading cell with windows (depending on the spectral range either CaF$_2$ or KBr) has been used to expose the thin film to hydrogen pressures up to $1.0\cdot10^5$ Pa at 298\,K. After hydrogenation for several hours at these conditions, transmission and reflection have been recorded from the substrate side. In-plane translation of the sample allowed scanning along the composition gradient.

\begin{figure}
\begin{center}
\includegraphics [width=\linewidth]{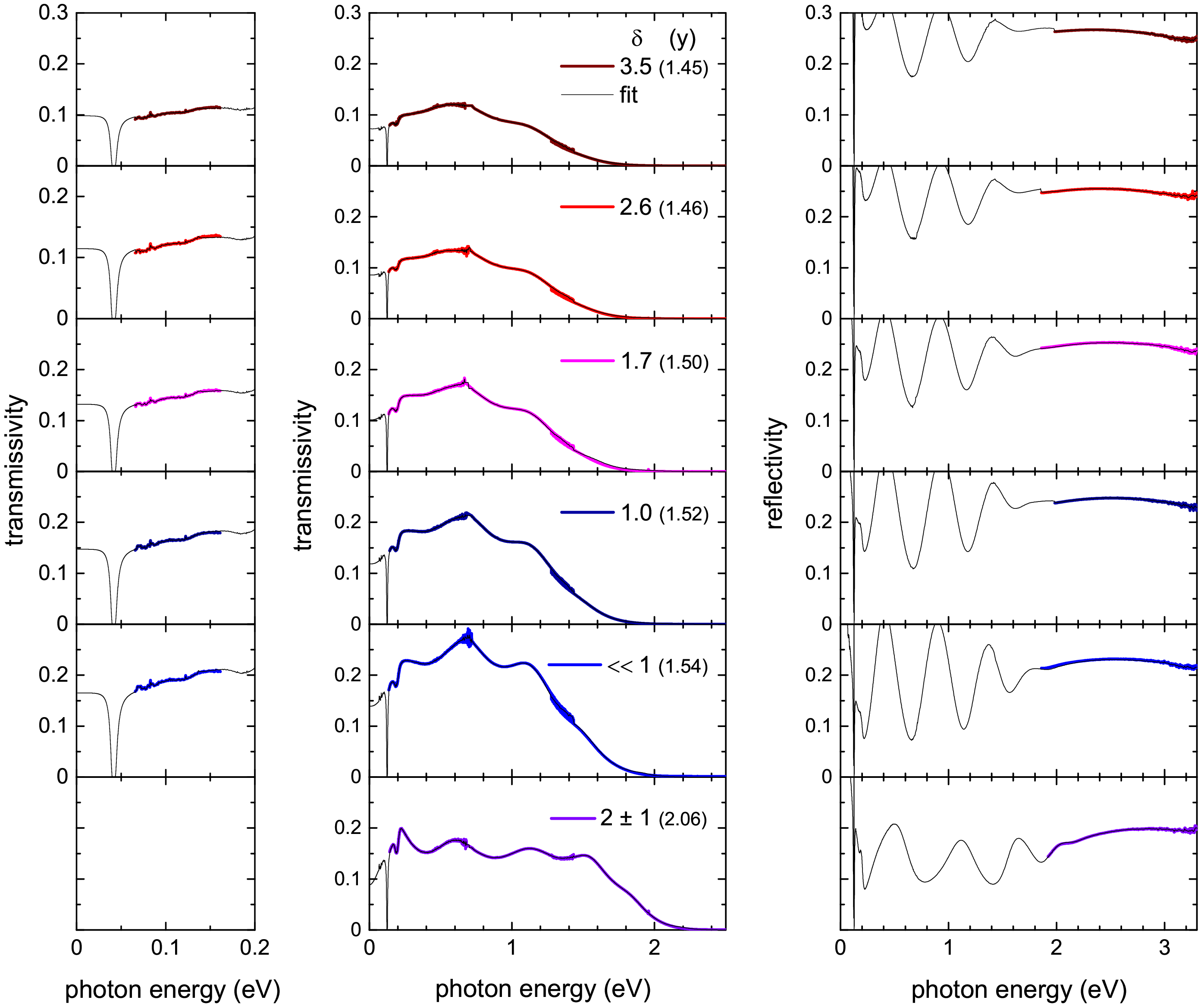}
\caption{Transmissivity $\mathcal{T}$ and reflectivity $\mathcal{R}$ of substrate -- Mg$_y$NiH$_{4-\delta/8}$ -- PdH$_z$ for a selection of the samples used in this work, characterised by $\delta$ ($y$ in parentheses) at 298\,K. The substrate is KBr for $\mathcal{T}$ in the range $0.06\leq\hbar\omega\leq0.16$\,eV and CaF$_2$ in the range $0.14\leq\hbar\omega\leq2.5$\,eV. The thickness of Mg$_y$NiH$_{4-\delta/8}$ on CaF$_2$ is 302 nm (3.5), 316 nm (2.6), 349 nm (1.7), 331 nm (1.0), 389 nm ($\ll 1$) and 382 nm ($2\pm1$) where $\delta$ is indicated in parentheses. Black lines show best fits using a variational dielectric function as described in the text.} \label{fig:opticfit}
\end{center}
\end{figure}

\subsection{Determination of $\sigma_1(\omega)$}\label{sec:fitting}
The measured reflectivity $\mathcal{R}$ and transmissivity $\mathcal{T}$ of CaF$_2$/KBr -- Mg$_y$NiH$_{4-\delta/8}$ -- PdH$_z$ stacks exposed to $1.0\cdot10^5$ Pa H$_2$ at 298\,K are shown in Fig.~\ref{fig:opticfit}. For all samples, Fabry-P\'erot interference within the Mg$_y$NiH$_{4-\delta/8}$ layer causes an oscillatory behavior, mainly visible in $\mathcal{T}$. The transmissivity shows infrared-active vibrational modes as dips around 0.08\,eV (Ni--H bending), 0.12\,eV (Mg--H) and 0.19\,eV (Ni--H stretching), in agreement with Refs.~\onlinecite{Richardson2001,Parker2002}.

The dielectric function $\varepsilon(\omega)=\varepsilon_1+(4\pi/\omega)i\sigma_1$ of Mg$_y$NiH$_{4-\delta/8}$ has been obtained by simultaneously fitting the transfer functions $\mathcal{R}$ and $\mathcal{T}$ for each value of $y$ based on the Fresnel equations for a stratified system using the Geneva package RefFit (A.B. Kuzmenko, University of Geneva).\cite{Kuzmenko2005, VanMechelen_OL2014} The sample thickness is determined from the thickness in the as-deposited state multiplied by a 15\,\% expansion due to hydrogenation.\cite{Lohstroh2004} This value been used as initial thickness value and is further optimized in the fitting procedure. Initial fitting has been done using a Drude-Lorentz parametrization of $\varepsilon(\omega)$. Oscillators were created for the following excitations: one for the coherent Drude peak, one for each vibrational mode, one for the infrared absorption band, and one for the original Mg$_2$NiH$_4$ interband transition around 1.7\,eV. The fitting parameters for each oscillator are the frequency $\omega_0$, the plasma frequency $\omega_p$ and the scattering rate $\gamma$.

\begin{figure}
\begin{center}
\includegraphics [width=0.8\linewidth]{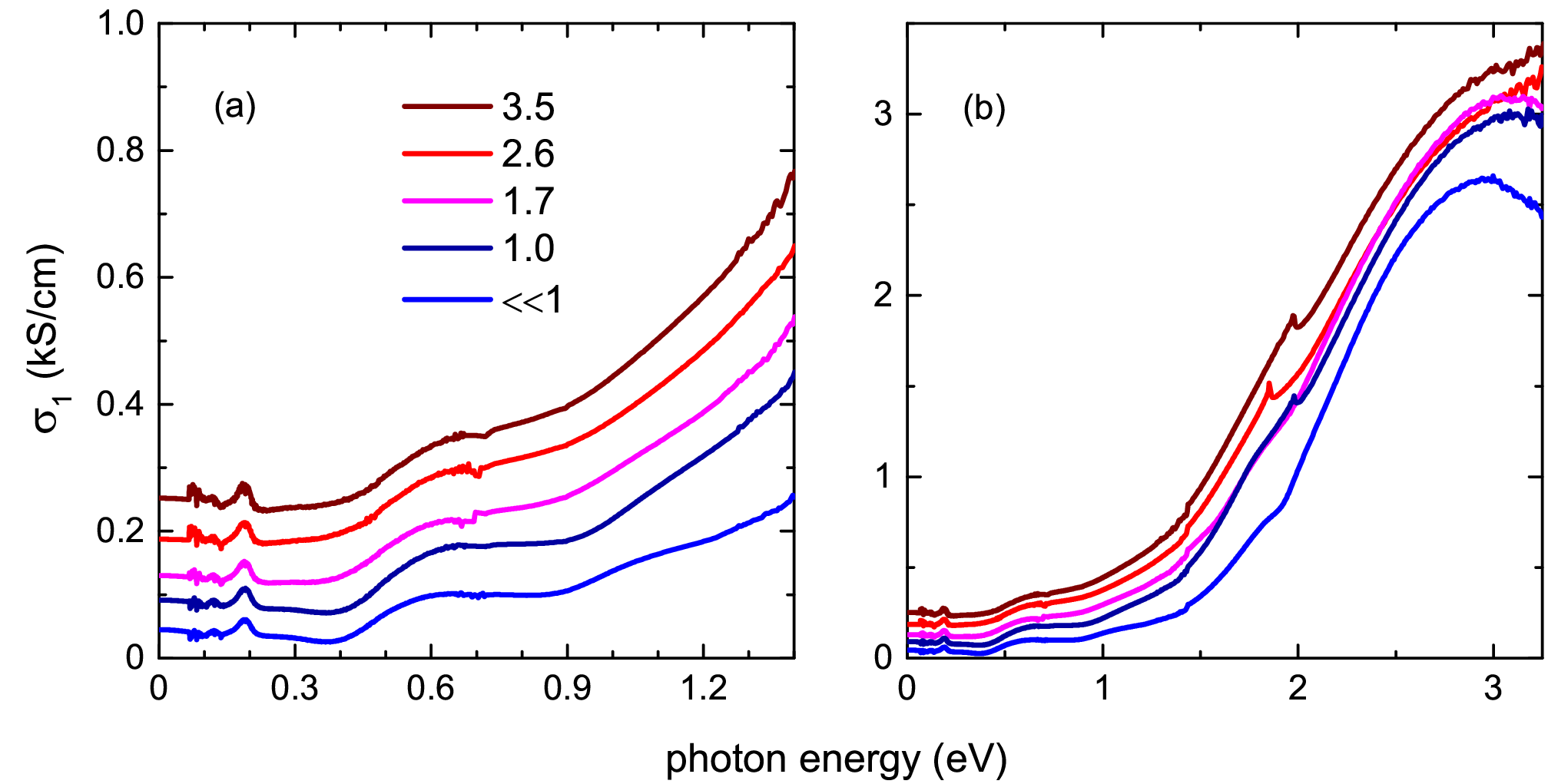}
\caption{Experimental $\sigma_1(\omega)$ of Mg$_y$NiH$_{4-\delta/8}$ at indicated values of $\delta$ obtained from using a variational dielectric function fitting procedure (see text). Around $\hbar\omega=0.7$ eV, 1.5 eV and 2 eV separately measured spectra join, and their mutual differences cause some irregularities in $\sigma_1$.} \label{fig:s1}
\end{center}
\end{figure}

A closer match to the natural dispersion has be obtained upon subsequent fitting with a variational dielectric function that enables a quasi line-shape independent fit.\cite{Kuzmenko2005, VanMechelen_OPN2015} Fig.~\ref{fig:opticfit} shows the fitted model transfer functions and Fig.~\ref{fig:s1} the resulting $\sigma_1(\omega)$. Note that the infrared regions for which the modeled $\mathcal{T}\rightarrow0$ correspond to the Reststrahlen bands of the KBr and CaF$_2$ substrates. As compared to a conventional Drude-Lorentz parameterization, the use of a variational dielectric function is of value for a precise evaluation of the dispersion and spectral weight.

\begin{figure}
\begin{center}
\includegraphics [width=0.5\linewidth]{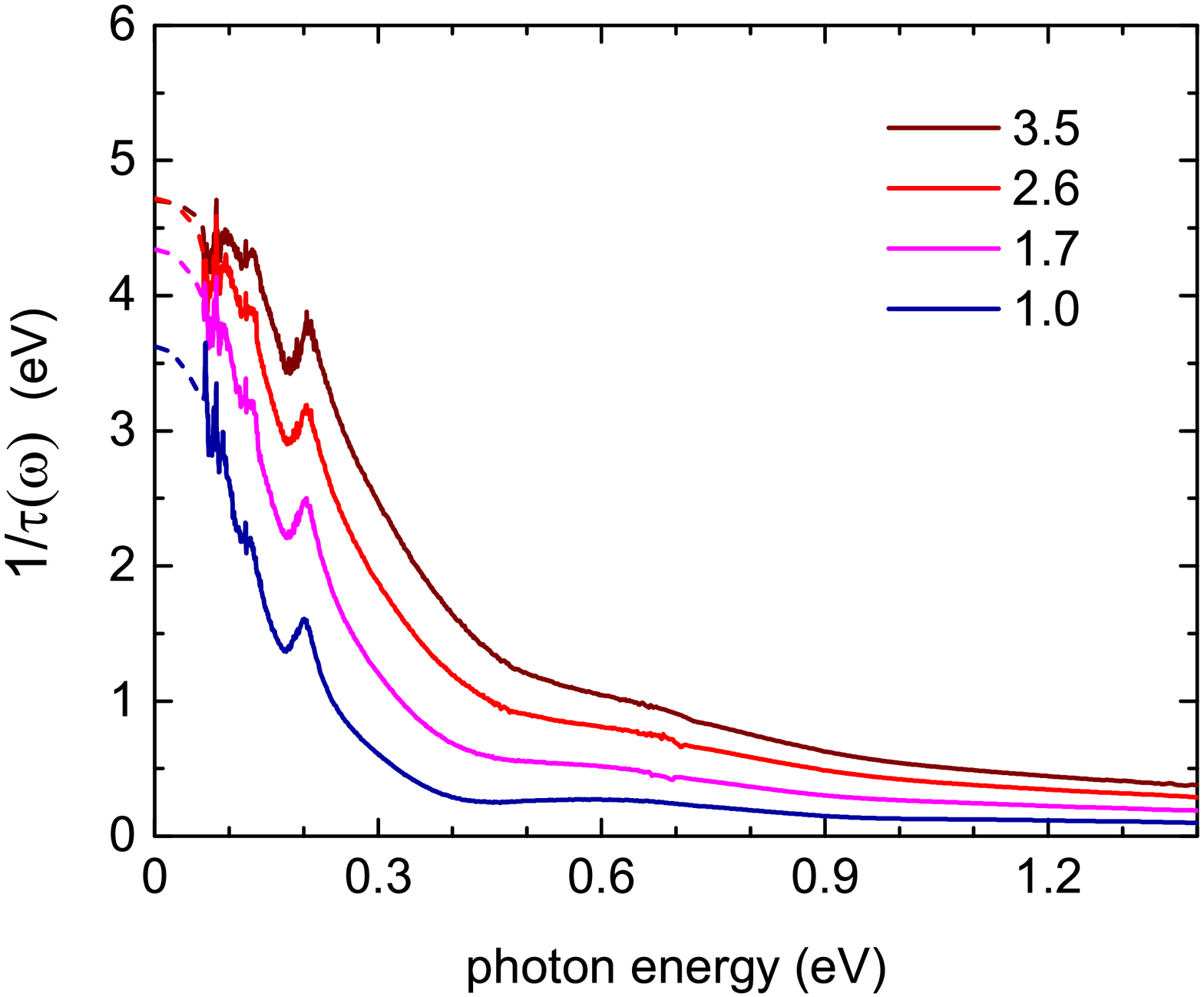}
\caption{$1/\tau(\omega)$ obtained from an extended Drude analysis for indicated values of $\delta$.} \label{fig:invtau}
\end{center}
\end{figure}

\subsection{Extended Drude analysis}\label{extendeddrude}
In systems with renormalized electron behavior, the conductivity is described by the extended Drude model, which takes into account the frequency dependence of the effective electron mass $m^*(\omega)$ and of the scattering rates $\tau(\omega)$ and $\tau^*(\omega) $,\cite{DresselGruner}
\begin{equation}\label{extendeddrude}
\frac{m^*(\omega)}{m}=\frac{\omega_{p}^2}{4\pi\omega}\text{Im}\left(\frac{1}{\sigma(\omega)}\right),\qquad \frac{1}{\tau(\omega)}=\frac{\omega_{p}^2}{4\pi}\text{Re}\left(\frac{1}{\sigma(\omega)}\right) \quad\text{and} \quad \frac{1}{\tau^*(\omega)}=\frac{1/\tau(\omega)}{m^*(\omega)/m},
\end{equation}
where $\omega_p$ is determined using the Hall carrier density (see section \ref{sec:hall}).

In the present case, the unrenormalized $1/\tau(\omega)$ is large at $\omega=0$ because $\omega_p$ and $\gamma$ of the coherent Drude peak have values in the same ballpark  (Fig.~\ref{fig:invtau}). Furthermore, $1/\tau(\omega)$ decreases with increasing photon energy due to the semiconductor nature of the material, which is characterized by the onset of very strong interband transitions above 1.7 eV. Concomitantly, $m^*(\omega)/m=\tau^*/\tau$ is negative as shown in Fig. 1a.

\subsection{Spectral weight transfer analysis}\label{optmass}
The electronic mass renormalization $m^*/m_e$ can be determined solely from optical data using a spectral weight transfer analysis. $m^*/m_e=W_{\text{tot}}/{W_{\text{coh}}}$, where the total electronic spectral weight $W_{\text{tot}}=W_{\text{coh}}+W_{\text{incoh}}$, being the sum of the coherent and incoherent Drude weights, is obtained from $\sigma_1(\omega)$ using the partial $f$-sum rule,

\begin{equation}\label{eq:fsumrule}
W_{\text{tot}}=\int_0^{\omega_c}\sigma_1(\omega)\,d\omega,
\end{equation}
where $\omega_c$ is a cut-off frequency. $\omega_c$ is to be chosen above the intraband transitions, that is above the coherent and incoherent Drude weights, and below the interband transitions.

\begin{figure}
\begin{center}
\includegraphics [width=0.75\linewidth]{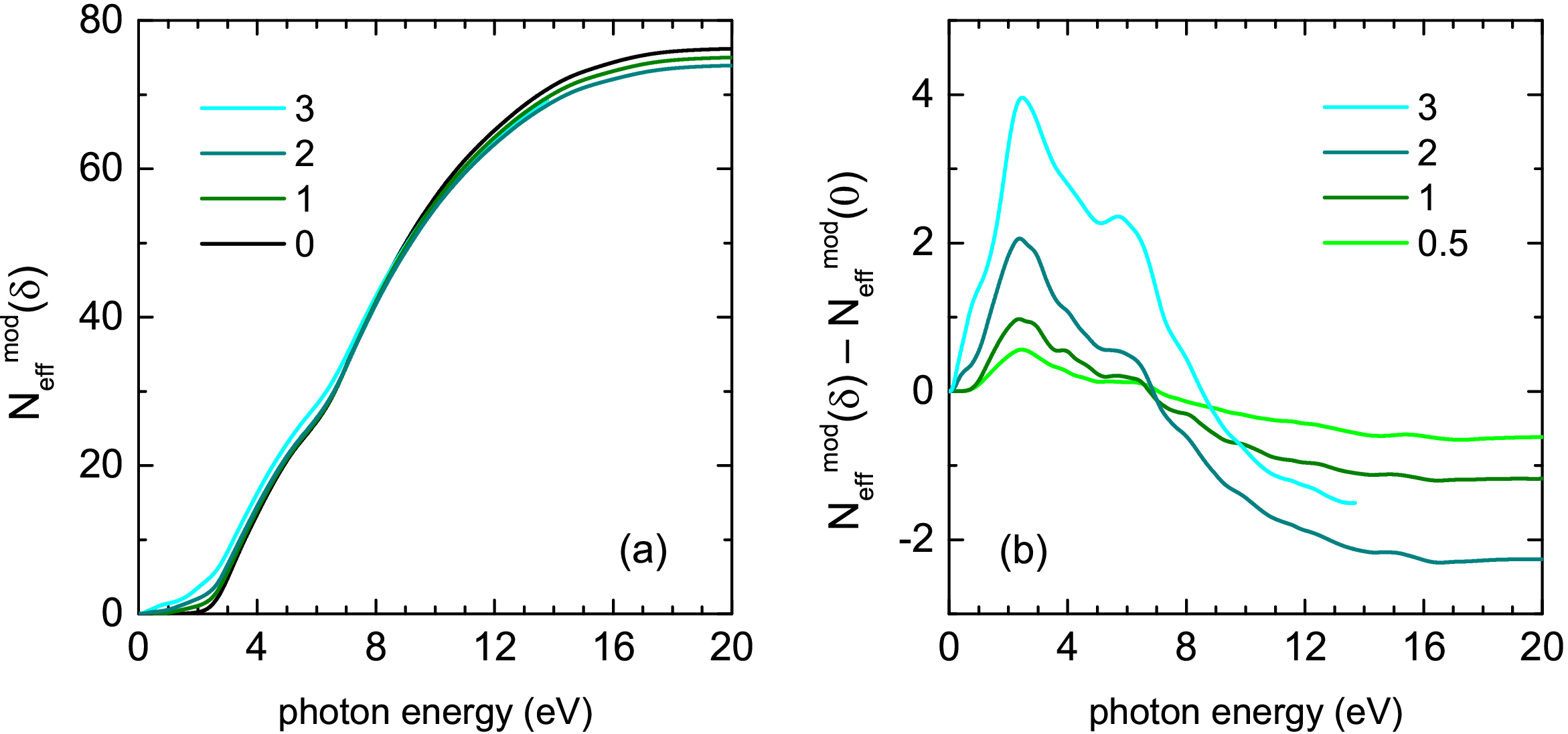}
\caption{(a) Effective carrier density $N_{\text{eff}}^{\text{mod}}$ determined from the calculated $\sigma_1(\omega)$, without the coherent Drude weight, for indicated values of $\delta$, (b) $N_{\text{eff}}^{\text{mod}}(\delta)-N_{\text{eff}}^{\text{mod}}(0)$.} \label{fig:neff}
\end{center}
\end{figure}

We determine $\omega_c$ based on the behavior of the effective carrier density $N_{\text{eff}}^{\text{mod}}=cW_{\text{tot}}V$, where $c=4.2568\cdot10^{14}$ and V the unit cell volume. $W_{\text{tot}}$ is determined from the calculated $\sigma_1(\omega)$, without the coherent Drude weight, using DFT (see section \ref{sec:dft}). Fig.~\ref{fig:neff}a shows $N_{\text{eff}}^{\text{mod}}$ for several charge carrier levels $\delta$ due to removal of H from the Mg$_2$NiH$_4$ unit cell. A clearer picture of the charge dynamics upon H removal is obtained by subtracting the interband transitions, manifested by $N_{\text{eff}}^{\text{mod}}(0)$, from $N_{\text{eff}}^{\text{mod}}(\delta)$ (Fig.~\ref{fig:neff}b). This is, however, only permitted for $\delta\leq3$, as the interband transitions for $\delta=4$ are significantly different from those at $\delta=0$ (cf.~inset of Fig.~3c). Removal of H obviously causes a spectral weight transfer across $\hbar\omega\approx10$\,eV to lower energies, also indicated by the DOS and $\sigma_1(\omega)$ (Figs.~2 and 3). Besides, the calculations show through $N_{\text{eff}}^{\text{mod}}(\delta)-N_{\text{eff}}^{\text{mod}}(0)$ at high energies, i.e., about 20 eV, that removal of each H atom indeed adds one charge carrier to the system, as reported experimentally.\cite{Enache2004} Fig.~\ref{fig:neff}b shows that $\omega_c$ is best determined to be around 2.5 eV where $N_{\text{eff}}^{\text{mod}}(\delta)-N_{\text{eff}}^{\text{mod}}(0)$ has a maximum which is very close to $\delta$ for $\delta<3$, indicating that the complete spectral weight corresponding to the charge carrier density $\delta$ is recovered at this photon energy.

Following the described procedure, we determine the optical mass renormalization $m^*/m_e$ (see Fig.~1). Hereto, $W_{\text{tot}}$ is obtained using Eq.~\ref{eq:fsumrule} with $\hbar\omega_c=2.5$\,eV. The vibrational spectral weight, being much smaller than the electronic spectral weight, has been subtracted after integration of $\sigma_1(\omega)$. $W_{\text{coh}}=(\pi/120)\omega_p^2$, where $\omega_p$ is obtained from the Drude-Lorentz fitting procedure.

\section{Charge carrier density from Hall data}\label{sec:hall}
The temperature dependence of the DC resistivity $\rho_{\text{DC}}$ and charge carrier density $n$ of Mg$_{2.17}$NiH$_x$ for $0\leq x\leq3.95$ has been reported for $4.2\leq T\leq295$\,K by Enache \em et al.\em\cite{Enache2004}  The authors experimentally show that the relationship between $n$ and $x$ is linear, independent of temperature.

In the present work, the charge carrier density per unit cell $\delta=n\cdot V$, where $V$ is the volume of the Mg$_2$NiH$_4$ unit cell (Table \ref{struc}). For each sample, $n$ is determined by mutual comparison of $1/\sigma_1$ at $\omega=0$ determined from the fitting procedure (see section \ref{sec:fitting}) and $\rho_{\text{DC}}(n)$ reported by Enache \em et al.\em\ This is justified since (i) $\rho_{\text{DC}}(n)$ is a monotonic function where $\rho_{\text{DC}}$ is uniquely determined by $n$,\cite{Enache2004} (ii) The role of Mg is only to structurally stabilize the Mg$_y$NiH$_x$ hydride and small variations of the Mg content $y$ are thus irrelevant to $\rho_{\text{DC}}(n)$, as shown by $\rho_{\text{DC}}(y)$ of Mg$_y$NiH$_x$ which is independent of $1.4\leq y\leq 2.5$,\cite{Mongstad2012} (iii) DFT calculations confirm that the electronic structure of Mg$_y$NiH$_x$ is essentially determined by $x$ and  invariant upon changing $y\approx2$ by 10-20\,\% (see section \ref{sec:DFT_y_delta}), and (iv) the sample batches of this work and those of Enache \em et al.\em\ are grown by the same technique in the same laboratory (VU University, Amsterdam) in the same time period, and have a similar structure.  In the as-deposited state at 295\,K, Enache \em et al.\em\ have $\rho_{\text{DC}}=55\,\mu\Omega$cm, the sample with the same composition of this work gives $\rho_{\text{DC}}=45\,\mu\Omega$cm, which is very close. In section \ref{sec:error} we discuss the expected error on $m^*/m_e$ for the given way of the determination of $\delta$.

\section{Uncertainty determination of $m^*/m$}\label{sec:error}
In the present study, an uncertainty on the determination of $\delta$ also impacts the uncertainty on $m^*/m_e=W_{\text{tot}}/{W_{\text{coh}}}=\omega^2_{p,\text{Hall}}/\omega^2_{p,\text{Drude}}$ since $\omega^2_{p,\text{Hall}}=4\pi\delta e^2/m_eV$. As described before, $\delta$ is obtained by comparison of $1/\sigma_1(0)$ to $\rho_{\text{DC}}(n)$ (section \ref{sec:hall}). The error determining parameters are thus $\sigma_1(0)$ and $\omega^2_{p,\text{Drude}}$. In order to estimate the variation of these parameters, we have performed multiple fitting procedures for each sample, thereby varying the individual weights of the used data sets. The resulting variation of $\sigma_1(0)$ and $\omega^2_{p,\text{Drude}}$ is then used to determine the error margins on $\delta$ and $m^*/m_e$ as shown in Fig.~1d.

\section{Density functional theory calculations}\label{sec:dft}

\subsection{Computational methods}
First principles density functional theory (DFT) calculations were carried out using a plane wave basis set and the projector augmented wave (PAW) method,\cite{paw,blo} as incorporated in the Vienna \em Ab initio \em Simulation Package (VASP).\cite{vasp1,vasp2,vasp3}
We use the PW91 generalized gradient approximation (GGA) for the exchange correlation functional.\cite{gga} Non-linear core corrections were applied.\cite{core}

The total energies were calculated on a $5\times5\times5$ $\mathbf{k}$-point mesh using a 700~eV kinetic energy cutoff. Since the zero point energies of MgH$_2$ (0.40 eV/H$_2$, Ref.~\onlinecite{VanSetten2005}) and Mg$_2$NiH$_4$ (0.42 eV/H$_2$, Ref.~\onlinecite{VanSetten2007}) are very close, we estimate that the zero point energy correction to the hydrogen desorption enthalpies of non-stoichiometric Mg$_2$NiH$_4$ will be almost constant (about 0.11~eV/H$_2$). We therefore did not calculate them explicitly for these structures.

The calculations of the interband dielectric functions were performed in the random phase independent particle approximation taking into account only direct transitions from occupied to unoccupied Kohn-Sham orbitals. No excitonic, local field, or $GW$ corrections have been taken into account. The imaginary part of the macroscopic dielectric function then has the form
\begin{equation}\label{epsinter}
\varepsilon^{\text{inter}}_2(\mathbf{\hat{q}},\omega)=\frac{8\pi^2 e^2}{V}\limq \frac{1}{\abs{\bq}^2}\sum_{\bk,v,c}\abs{\bracet{u_{c,\bk + \bq}}{u_{v,\bk}}}^2
\delta(\epsilon_{c,\bk + \bq}-\epsilon_{v,\bk}-\hbar\omega)
\end{equation}
where $\mathbf{\hat{q}}$ gives the direction of $\bq$ and $v,\mathbf{k}$ and $c,\mathbf{k}$ label single particle states that are occupied, unoccupied in the ground state, respectively, and $\epsilon$, $u$ are the single particle energies and the translationally invariant parts of the wave functions; $V$ is the volume of the unit cell. The real part of $\varepsilon$ is obtained via a Kramers-Kronig transformation. Further details on the calculation of the optical properties can be found in Ref.~\onlinecite{kresseps}. A $7\times7\times5$ Monkhorst-Pack $\mathbf{k}$-point mesh was used for all the optical calculations.\cite{monk} For the larger super-cells the $\mathbf{k}$-point mesh was reduced accordingly.

Almost all optical data on hydrides are obtained from micro-crystalline samples whose crystallites have a significant spread in orientation. The most relevant quantity then is the directionally averaged dielectric function, i.e., $\varepsilon_{2}(\omega)$ averaged over $\mathbf{\hat{q}}$.

Starting from the experimental positions and lattice parameters, as reported in Ref.~\onlinecite{zolliker86}, we relaxed the crystal structure of Mg$_2$NiH$_4$. At a range of volumes we relaxed all lattice parameters and atomic positions, thereby minimizing the total energy of the electron-ion system. To the energy vs.\ volume data obtained in this way we fitted the Murnaghans equation of state obtaining the ground state volume and bulk modulus. We calculate an optimal unit cell volume of 546~\AA$^3$ (68.3~\AA$^3$/f.u.) and a bulk modulus of 44~GPa. H\"aussermann \em et al.\em\ calculated an equilibrium volume of 68.44~\AA$^3$/f.u.\ and a bulk modulus of 50~GPa.\cite{haussermann02} At the obtained volume, the lattice parameters and atomic positions were relaxed once more. The results, given in Table~\ref{struc}, agree well with experimental\cite{zolliker86} and previously calculated\cite{haussermann02,myers02} values. Note that in Ref.~\onlinecite{myers02} the lattice parameters were kept fixed at the experimental values.

\begin{table}
\caption{\label{struc}Calculated atomic positions and lattice parameters of Mg$_2$NiH$_4$, space group C2/c (15).}
\begin{ruledtabular}
\begin{tabular}{lcccccc}
 unit cell & & & x & y & z \\
\hline
$\beta$ = 113.35$^\circ$ & Mg &$8f$& 0.2646 & 0.4871 & 0.0835 \\
a = 14.37~\AA            & Mg &$4e$& 0      & 0.0263 & 0.2500 \\
b = 6.39~\AA             & Mg &$4e$& 0      & 0.5279 & 0.2500 \\
c = 6.48~\AA             & Ni &$8f$& 0.1202 & 0.2294 & 0.0794 \\							
                         & H  &$8f$& 0.2096 & 0.3039 & 0.3033 \\
                         & H  &$8f$& 0.1390 & 0.3213 & 0.8760 \\
                         & H  &$8f$& 0.0110 & 0.2920 & 0.0559 \\
                         & H  &$8f$& 0.1243 & 0.9865 & 0.0715 \\
\end{tabular}
\end{ruledtabular}
\end{table}

\subsection{Mg$_y$NiH$_{x-\delta/8}$: controlling $y$ at $\delta=0$}\label{sec:dft_y}

From the relaxed structure we construct the non-stoichiometric models. We start with a conventional unit cell (Z\,=\,8) and remove nickel atoms. Subsequently we relax the atomic positions and lattice parameters. The magnesium atoms are then expected to donate two electrons to either, single hydrogen atoms, which then form a two electron 1$s$ closed shell (as in MgH$_2$) or to a NiH$_4$ complex, which then forms an eighteen electron closed shell configuration (as in Mg$_2$NiH$_4$). An inspection of the densities of states of the non-stoichiometric models confirms this picture. Because the Mg$_y$NiH$_4$ models are fully hydrogenated, all electrons are in closed shell configurations and  all models become semiconductors.

Based on calculated and experimental spectra, we here demonstrate that variation of $y$ at complete hydrogenation mainly changes the optical properties above the Mg$_2$NiH$_4$ band gap.

\begin{figure}[b]
\begin{center}
\includegraphics[width=0.5\linewidth]{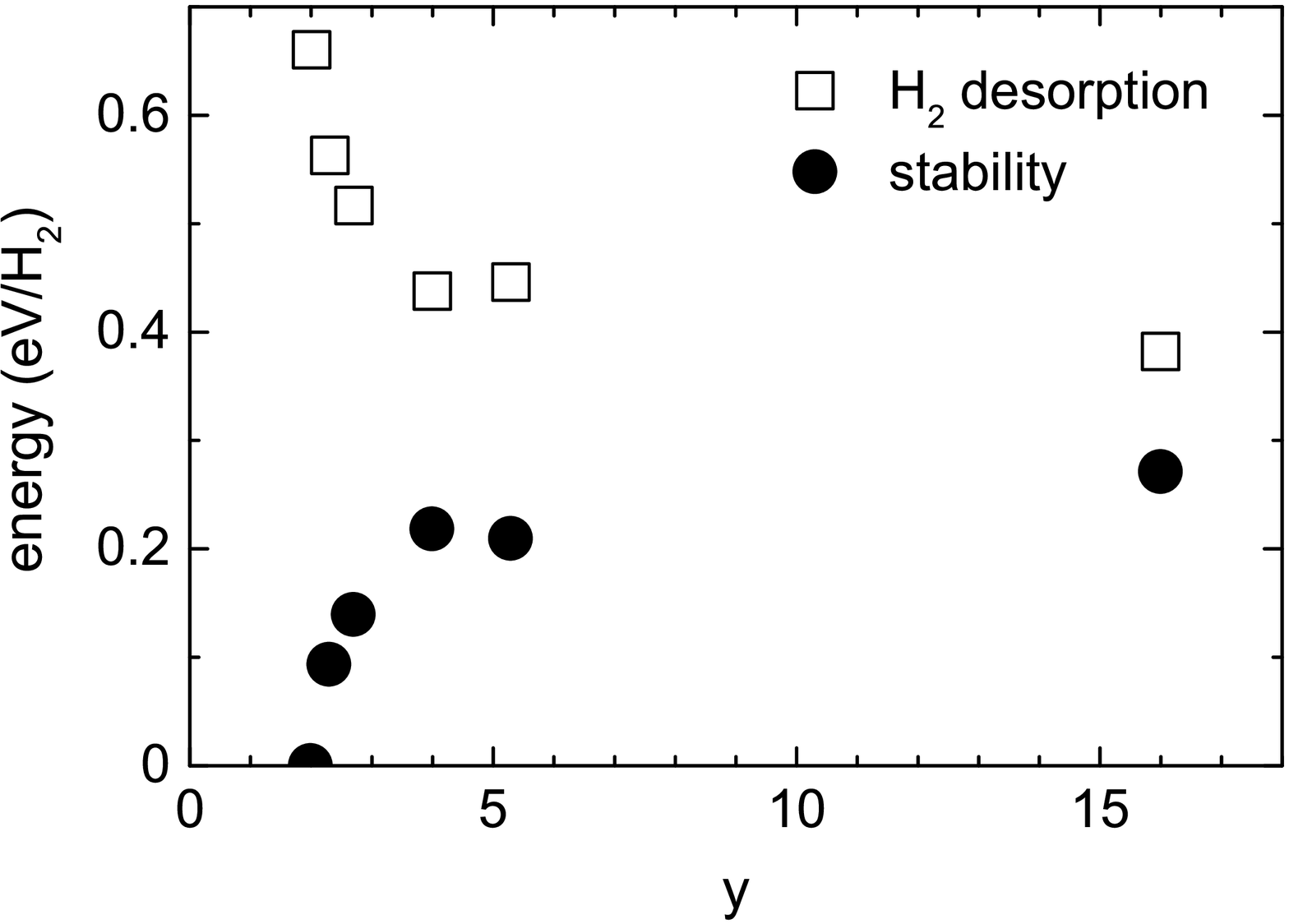}
\caption{\label{desorp}Calculated hydrogen desorption enthalpy (with respect to Mg+Mg$_2$Ni+H$_2$) and stability with respect to phase segregation (into Mg$_2$NiH$_4$ and MgH$_2$) of Mg$_y$NiH$_4$.}
\end{center}
\end{figure}

We calculate the hydrogen desorption enthalpy and the stability of the non-stoichiometric model systems using their total energies and the total energies of bulk Mg, MgH$_2$, Mg$_2$Ni, and H$_2$ as reported earlier.\cite{VanSetten2007} The hydrogen desorption enthalpy is calculated with respect to Mg$_2$Ni and bulk Mg and the stability with respect to phase segregation into Mg$_2$NiH$_4$ and MgH$_2$. From Fig.~\ref{desorp} it is clear that all models for the non-stoichiometric systems are metastable and that the hydrogen desorption enthalpy is lower than that of Mg$_2$NiH$_4$.

For stoichiometric Mg$_2$NiH$_4$ we calculate a band gap of 1.65~eV which is remarkably close to the experimental value of 1.68~eV.\cite{lupu87} This close resemblance is rather unique for a DFT calculation, which usually underestimates the band gap on the order of one eV. We conjecture that errors introduced by using the DFT band structure, which usually underestimates the band gap, cancel errors introduced by employing the RPA approximation. The latter neglects exciton effects and hence overestimates the optical gap. Full $GW$+BSE calculations, which would remedy both of these approximations, are however beyond the scope of this work. Especially the complex defect super cells computationally prohibit such an approach. Previous calculations have shown gaps of 1.54 and 1.4~eV.\cite{haussermann02,myers02}

\begin{figure}
\begin{center}
\includegraphics [width=0.5\linewidth]{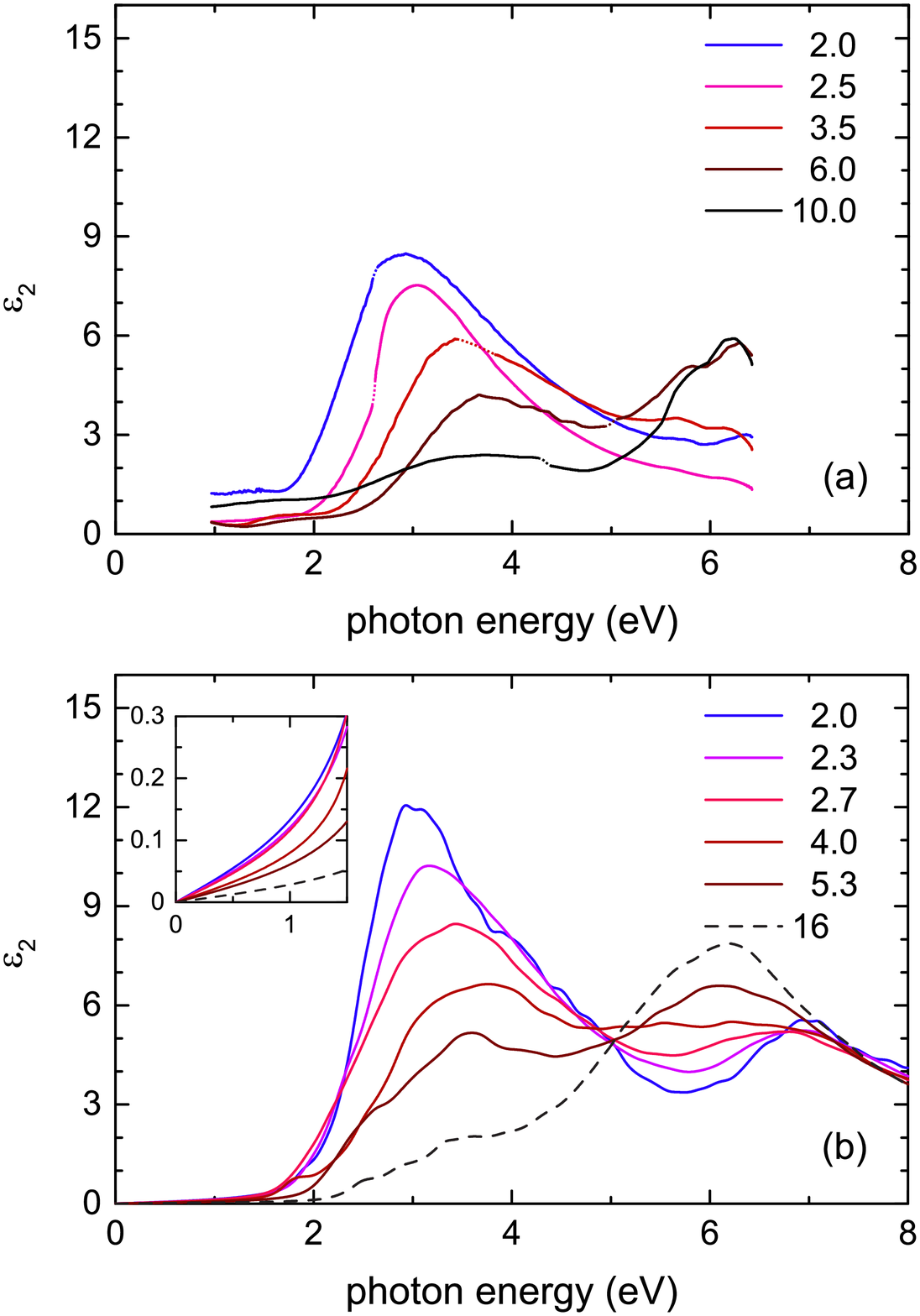}
\caption{$\varepsilon_2(\omega)$ of Mg$_y$NiH$_x$ in the visible and ultraviolet spectral range for values of $y$ indicated in the legends. (a) Using experimental data from Ref.~\onlinecite{Lohstroh2006}.\cite{note1} (b) Calculated for non-stoichiometric models Mg$_y$NiH$_4$. The inset of (b) shows the calculated $\varepsilon_2(\omega)$ for $0\leq \hbar\omega \leq 1.5$ eV.} \label{fig:e2-visuv}
\end{center}
\end{figure}

Fig.~\ref{fig:e2-visuv}(b) displays the calculated $\varepsilon_2(\omega)$ of non-stoichiometric Mg$_y$NiH$_4$. Upon increasing $y$, the characteristics of $\varepsilon_2(\omega)$ gradually change from Mg$_2$NiH$_4$ to MgH$_2$. Similarity with the experimentally determined $\varepsilon_2(\omega)$ can be seen by the band edge variation for $y$ between 2.5 and 6, by the composition behavior of the 2-4~eV absorption band, and for $y>5$ by the upcoming band edge of MgH$_2$ around 6~eV (cf.~Fig.~\ref{fig:e2-visuv}a)~\cite{note2}.

The models show that $\epsilon_2(\omega)$ in the infrared is small and mainly determined by the roll-off of interband transitions across the Mg$_2$NiH$_4$ band gap, with a minor dependence on $y$. The experimental $\epsilon_2(\omega)$ in the infrared is much larger than the models predict (see the main text of this work and Ref.~\onlinecite{Lohstroh2006}) due to incomplete hydrogenation as discussed in the section below.

\subsection{Mg$_y$NiH$_{x-\delta/8}$: controlling $\delta$ at $y=2$}

We study the influence of incomplete hydrogenation using DFT. At first, we consider four different vacancy concentrations in the calculations: one neutral hydrogen atom removed from a $2\times 2\times 1$, a $2\times 1\times 1$ and a $1\times 1\times 1$ conventional super-cell and two hydrogen atoms removed from a $1\times1\times1$  cell. For every case, all atomic positions are optimized.

Fig.~2c shows the electronic densities of states (DOS) of Mg$_2$NiH$_{x-\delta/8}$ with $\delta$ charge carriers per unit cell. For pristine Mg$_2$NiH$_4$ the DOS shows a fundamental gap of about 1.7 eV which separates the valence band of H $1s$ and Ni $3d$ character, and the conduction band of Mg $s$, $p$ and $d$ character. Upon removal of (neutral) hydrogen atoms in super-cells of decreasing size, i.e., with increasing $\delta$, states arise in the original gap of Mg$_2$NiH$_4$. Calculating the site projected densities of states within the PAW formalism indicates that all these in-gap states have rather mixed Mg, Ni and H character, independent of $\delta$.

When only one atom is removed from the different super-cells, the in-gap states have a single main peak, which originates from a single band. Within this peak no optical interband transitions are possible and the corresponding part of the dielectric function stays zero up to about 0.5~eV. From this energy onward, optical interband transitions are possible, which increases $\sigma_1=\omega\varepsilon_2/4\pi$, as shown in Fig.~3c. First excitations from the valence band to the unoccupied part of the in-gap states become accessible and from about 2~eV also excitations to the original conduction band states.

\begin{figure}
\begin{center}
\includegraphics [width=0.7\linewidth]{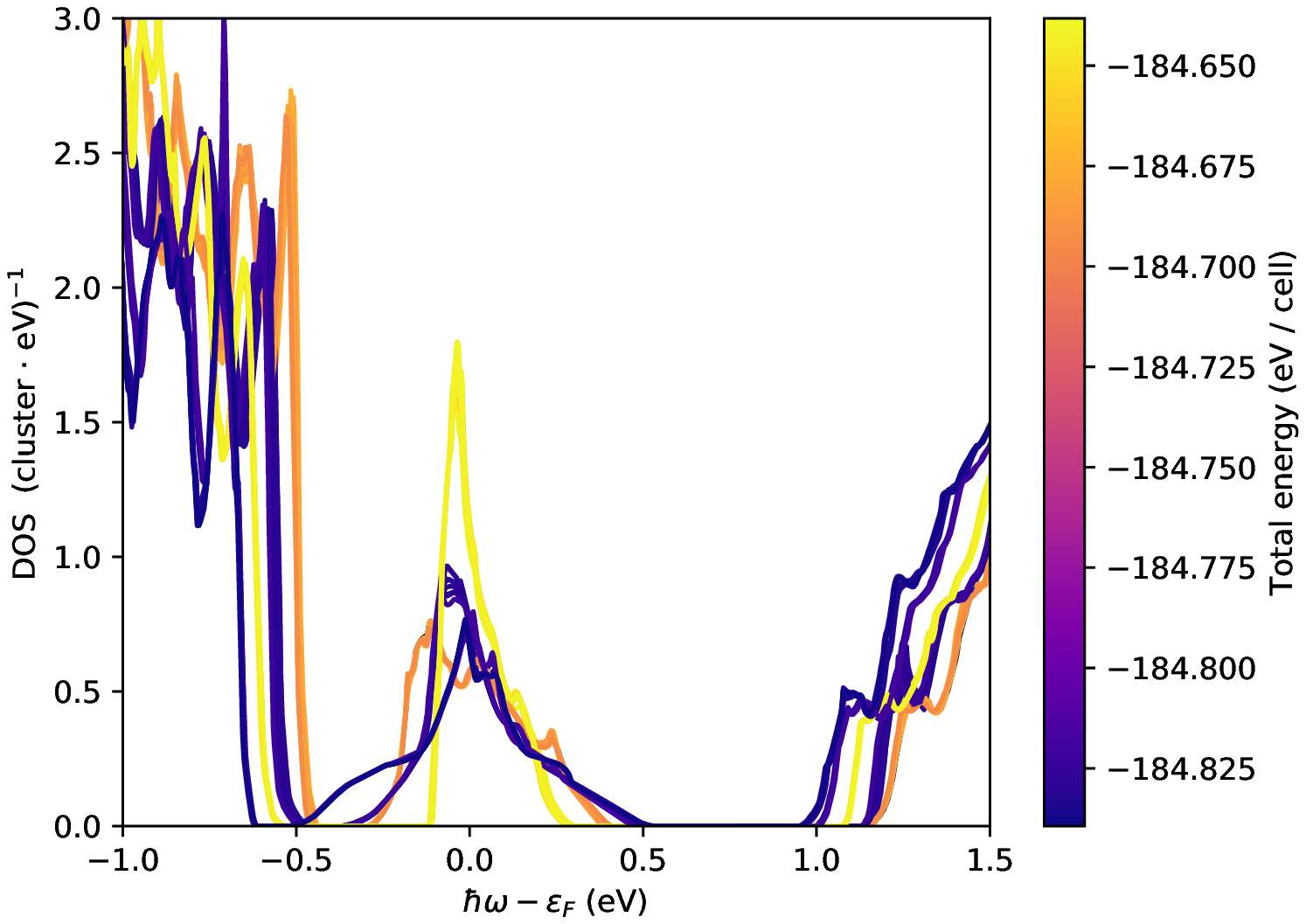}
\caption{DOS for super-cells of Mg$_2$NiH$_{4-\delta/8}$ with $\delta=1$, repeated for each hydrogen atom in the unit cell. The color of the line is given by the total energy of the optimized geometry, indicating the relative stability.} \label{fig:1on1-many-dos}
\end{center}
\end{figure}

To test the stability of the results against the defect position, we repeated the single hydrogen removal for {\em all} hydrogen atoms in the cell. Figure~\ref{fig:1on1-many-dos} shows the DOS for all these models. The color of each line indicates the stability of the system.

\begin{figure}
\begin{center}
\includegraphics [width=0.7\linewidth]{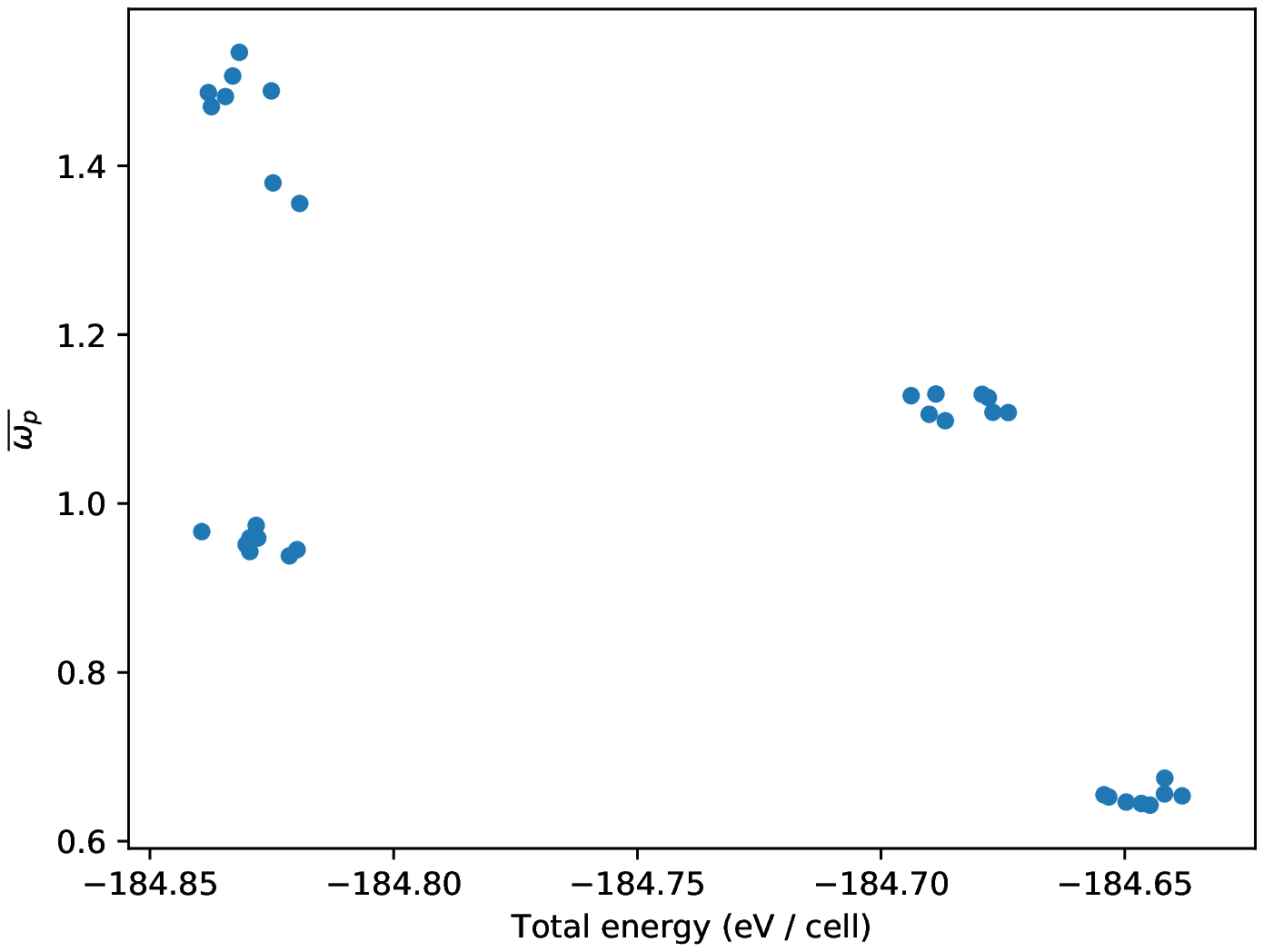}
\caption{The average of the diagonal elements of the intraband plasma frequency tensor plotted as a function of the total energy of the optimized cell for the 32 single hydrogen defect cells.} \label{fig:1on1-many-envswp}
\end{center}
\end{figure}

The intraband plasma frequencies, the average of the diagonal elements of the tensor, for the 32 single hydrogen defect cells are compared in Figure~\ref{fig:1on1-many-envswp}, where they are plotted as a function of the total energy of the cell. The four clusters, also visible in Figure~\ref{fig:1on1-many-dos}, originate from the four symmetry inequivalent hydrogen positions in the cell. However, once one hydrogen is removed all symmetry is broken and small differences in the structural optimization lead to the spread in the clusters.

From Fig.~\ref{fig:1on1-many-dos} and \ref{fig:1on1-many-envswp} we conclude that both the DOS of the single defect cell and its intraband plasma frequency are stable with respect to the position of the hydrogen defect. Especially no qualitative differences are observed in the DOS and the spread in the plasma frequencies does not influence the conclusions being drawn in the main text.

When two hydrogen atoms are removed from two different clusters, the in-gap states split into two peaks around $E_F$ (Fig. 2). Both are hybridized, such that the system is still metallic. Apparently, the two clusters are slightly different giving rise to shifted energy bands of which, in parts of the Brillouin zone, one is occupied and the other empty. Between these two levels low energy transitions are possible. Moreover, the defected clusters are now close enough to interact enabling the actual transitions. In the optical conductivity a strong new peak arises around 0.25~eV.

Analogously, we have calculated models with three, four and eight hydrogen atoms removed from one unit cell. Note that in the structure with $\delta=3$, two H atoms are removed from one plane and a third from a neighboring plane, giving rise to more H-like bands and thus a `blurred' DOS around $\varepsilon_F$ as compared to the DOS of $\delta=2$ and 4 where all H vacancies are in a single plane.

\begin{figure}
\begin{center}
\includegraphics [width=\linewidth]{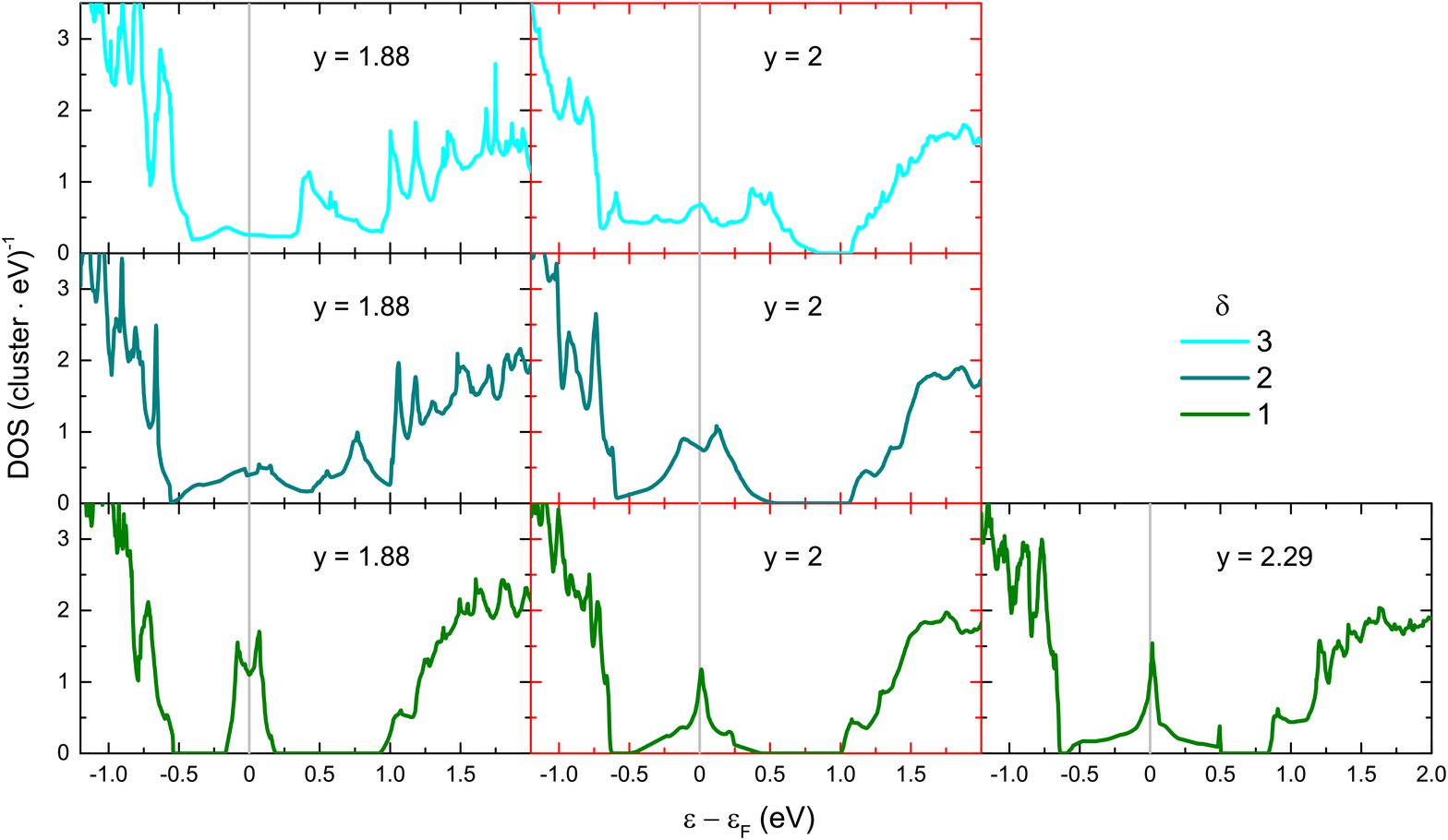}
\caption{Electronic density of states of Mg$_y$NiH$_{x-\delta/8}$ for $\delta=1, 2$ and 3 and selected values of $y$.} \label{fig:dos}
\end{center}
\end{figure}

\subsection{Mg$_y$NiH$_{x-\delta/8}$: controlling $y$ and $\delta$}\label{sec:DFT_y_delta}
Similar DFT calculations as discussed above are performed for the following optimized structures with varying both $y$ and $\delta$:\\ \\
\begin{itemize}
  \item Removal of one Ni atom from Mg$_2$NiH$_{x-\delta/8}$: Mg${_{16}}$Ni$_{7}$H$_{8x-\delta}$ ($y=2.29$);
  \item Removal of one Mg atom from Mg$_2$NiH$_{x-\delta/8}$: Mg${_{15}}$Ni$_{8}$H$_{8x-\delta}$ ($y=1.88$).
\end{itemize}

Fig.~\ref{fig:dos} shows the DOS for $y=2$ as compared to Mg-rich and Ni-rich cases. In all cases, the DOS is qualitatively similar to the DOS at $y=2$.